\documentclass[twocolumn]{emulateapj}

\newcommand{\LIR}{$L_{\rm IR}$}
\newcommand{\Msun}{$M_\odot$}
\newcommand{\Msunyr}{$M_\odot$yr$^{-1}$}
\newcommand{\Msunyrkpcsq}{$M_\odot$yr$^{-1}$kpc$^{-2}$}
\newcommand{\SFRSD}{$\Sigma_{\rm SFR}$}

\newcommand{\uJy}{$\mu$Jy}
\newcommand{\uJybm}{$\mu$Jy\,beam$^{-1}$} 
\newcommand{\logMstar}{log($M_*/M_{\odot}$)}

\begin{document}


\title{Cospatial Star Formation and Supermassive Black Hole Growth in $\lowercase{\it z} \sim 3$ Galaxies: Evidence for In-situ Co-evolution}


\author{W.~Rujopakarn\altaffilmark{1,2,3},
K.~Nyland\altaffilmark{4},
G. H.~Rieke\altaffilmark{5},
G.~Barro\altaffilmark{6},
D.~Elbaz\altaffilmark{7},
R. J.~Ivison\altaffilmark{8,9},
P.~Jagannathan\altaffilmark{10},\\
J. D.~Silverman\altaffilmark{3},
V.~Smol\v{c}i\'{c}\altaffilmark{11}, and 
T.~Wang\altaffilmark{7,12}
}

\affil{$^1$Department of Physics, Faculty of Science, Chulalongkorn University, 254 Phayathai Road, Pathumwan, Bangkok 10330, Thailand\\
$^2$National Astronomical Research Institute of Thailand (Public Organization), Don Kaeo, Mae Rim, Chiang Mai 50180, Thailand\\
$^3$Kavli Institute for the Physics and Mathematics of the Universe (WPI), The University of Tokyo Institutes for Advanced Study, \\ The University of Tokyo, Kashiwa, Chiba 277-8583, Japan; wiphu.rujopakarn@ipmu.jp\\
$^4$National Radio Astronomy Observatory, Charlottesville, VA 22903, USA\\
$^5$Steward Observatory, University of Arizona, Tucson, AZ 85721, USA\\
$^6$University of California, Berkeley, CA 94720, USA\\
$^{7}$CEA Saclay, DSM/Irfu/Service d'Astrophysique, Orme des Merisiers, F-91191 Gif-sur-Yvette Cedex, France\\
$^8$European Southern Observatory, Karl Schwarzschild Strasse 2, Garching, Germany\\
$^9$Institute for Astronomy, University of Edinburgh, Royal Observatory, Blackford Hill, Edinburgh EH9 3HJ, UK\\
$^{10}$National Radio Astronomy Observatory, Socorro, NM 87801, USA\\
$^{11}$Department of Physics, Faculty of Science, University of Zagreb,  Bijeni\v{c}ka cesta 32, 10000 Zagreb, Croatia\\
$^{12}$Institute of Astronomy, The University of Tokyo, 2-21-1 Osawa, Mitaka, Tokyo, 181-0015 Japan
}




\begin{abstract}

We present a sub-kpc localization of the sites of supermassive black hole (SMBH) growth in three active galactic nuclei (AGN) at $z \sim 3$ in relation to the regions of intense star formation in their hosts. These AGNs are selected from Karl G. Jansky Very Large Array (VLA) and Atacama Large Millimeter/submillimeter Array (ALMA) observations in the HUDF and COSMOS, with the centimetric radio emission tracing both star formation and AGN, and the sub/millimeter emission by dust tracing nearly pure star formation. We require radio emission to be $\geqslant5\times$ more luminous than the level associated with the sub/millimeter star formation to ensure that the radio emission is AGN-dominated, thereby allowing localization of the AGN and star formation independently. In all three galaxies, the AGN are located within the compact regions of gas-rich, heavily obscured, intense nuclear star formation, with $R_e = 0.4-1.1$ kpc and average star formation rates of $\simeq100-1200$ \Msunyr. If the current episode of star formation continues at such a rate over the stellar mass doubling time of their hosts, $\simeq 0.2$ Gyr, the newly formed stellar mass will be of the order of $10^{11}$ \Msun\ within the central kpc region, concurrently and cospatially with significant growth of the SMBH. This is consistent with a picture of {\it in-situ} galactic bulge and SMBH formation. This work demonstrates the unique complementarity of VLA and ALMA observations to unambiguously pinpoint the locations of AGN and star formation down to $\simeq30$ mas, corresponding to $\simeq 230$ pc at $z = 3$. 

\end{abstract}

\keywords{editorials, notices --- 
miscellaneous --- catalogs --- surveys}

\keywords{galaxies:evolution --- galaxies:star formation}



\section{Introduction} \label{sec:intro}

Multiple lines of evidence show a link between galaxy assembly and supermassive black hole (SMBH) growth, and that the accretion activity onto the SMBH leaves a lasting imprint on the evolution of its host galaxy \citep[][and references therein]{Kormendy2013, HeckmanBest14}. Over the past decade, there has been considerable effort to determine how well galaxies that harbor active galactic nuclei (AGN) are connected to the general galaxy population through large extragalactic surveys. A picture is emerging where there is a preference for AGN, with moderate to high accretion rates, to reside in star-forming galaxies once selection effects are under control. This is seen across redshifts from $z \lesssim 0.3$ \citep{Kauffmann03} up to $z \sim 1$ \citep{Silverman09} and beyond \citep{Mullaney12, Bongiorno16}. As a result, we have a clear association between accretion onto SMBHs and star formation, likely indicative of a co-evolution scenario on galaxy-wide scales.

However, a crucial aspect of improving our understanding of the link between SMBHs and star formation requires higher resolution imaging to isolate, within galaxies, whether SMBHs are associated with the sites where the bulk of star formation is occurring. Yet such images have been impossible at $z \sim 1-3$, potentially the formative era for the current relation. The commonly used tracers of AGN do not have the required resolution, e.g., the resolution in the most sensitive {\it Chandra} X-ray observations of the Chandra Deep-Field South \citep{Luo17} is typically $\simeq0\farcs7 - 3.6''$, corresponding to $5-27$ kpc at $z = 3$. In addition, the strong dust extinction typical in rapidly assembling star-forming galaxies (SFGs) at this epoch \citep[e.g.,][]{Dunlop17} requires an extinction-independent tracer of star formation such as far infrared, but again resolutions of $\sim 5''$, i.e., few tens of kpc, are typical. 

\begin{deluxetable*}{lccccccccccc}
\tablenum{1}
\tablecaption{The sample of radio-dominated AGNs with ALMA detection \label{tab:sample}}
\tablehead{\colhead{ID} & \colhead{RA} & \colhead{Dec} & \colhead{$z$} & \colhead{$M_*$} & \colhead{$S_{\rm VLA}$} & \colhead{$S_{\rm ALMA}$} & \colhead{$\nu_{\rm ALMA}$} & \colhead{log$L_{1.4}$} & \colhead{SFR} & \colhead{$M_{\rm gas}$} & \colhead{$f_{\rm gas}$}\\
\colhead{} & \colhead{(deg)} & \colhead{(deg)} & \colhead{} & \colhead{(log$M_{\odot}$)} & \colhead{(\uJy)} & \colhead{(\uJy)} & \colhead{(GHz)} & \colhead{(W\,Hz$^{-1}$)} & \colhead{(\Msunyr)} & \colhead{(log\Msun)} & \colhead{}
}
\startdata
UDF7 &  53.1805 & $-$27.7797 & 2.59 & $10.6 \pm 0.1$ & $18.7 \pm 0.6$ & $ 231 \pm  48$ & 232.8 & 24.3 & $ 107 \pm  12$ & $10.5^{+0.2}_{-0.2}$ & $0.42^{+0.12}_{-0.08}$\\
COS1 & 149.6886 &   2.2613 & 2.87 & $10.7 \pm 0.1$ & $  712 \pm 4$ & $4533 \pm 269$ & 336.5 & 25.7 & $570 \pm 76$ & $11.3^{+0.3}_{-0.2}$ & $0.81^{+0.08}_{-0.08}$\\
COS2 & 150.6694 &   2.1083 & 2.92 & $11.1 \pm 0.1$ & $  288 \pm 4$ & $4418 \pm 265$ & 336.5 & 25.4 & $1172 \pm 445$ & $11.3^{+0.3}_{-0.2}$ & $0.63^{+0.14}_{-0.11}$\\
COS4$^*$ & 149.8989 &   1.9682 & 3.25 & $10.9 \pm 0.3$ & $  263 \pm 3$ & $1101 \pm 115$ & 231.0 & 25.4 & \phantom{0}$492 \pm 118$ & $11.1^{+0.2}_{-0.2}$ & $0.59^{+0.12}_{-0.09}$\\
COS5$^*$ & 149.7473 &   1.7533 & 3.83 & $10.4 \pm 0.2$ & $  459 \pm 4$ & $ \phantom{0}505 \pm 109$ & 231.0 & 25.8 & \phantom{0}$583 \pm 198$ & $10.7^{+0.3}_{-0.2}$ & $0.67^{+0.12}_{-0.10}$\\
COS6$^*$ & 150.7434 &   2.1705 & 1.29 & $11.6 \pm 0.0$ & $  960 \pm 6$ & $1159 \pm 216$ & 336.5 & 25.1 & $444 \pm  65$ & $10.7^{+0.2}_{-0.1}$ & $0.12^{+0.06}_{-0.03}$
\enddata
\tablecomments{We assume $S_{\nu} \propto \nu^{-0.7}$ radio spectral slope to estimate the 1.4-GHz radio power, $L_{1.4}$. The asterisks in the ID column indicate objects in the supplementary sample (Section \ref{sec:obs}).}
\end{deluxetable*}

\begin{figure*}
\figurenum{1}
\centerline{\includegraphics[width=0.95\textwidth]{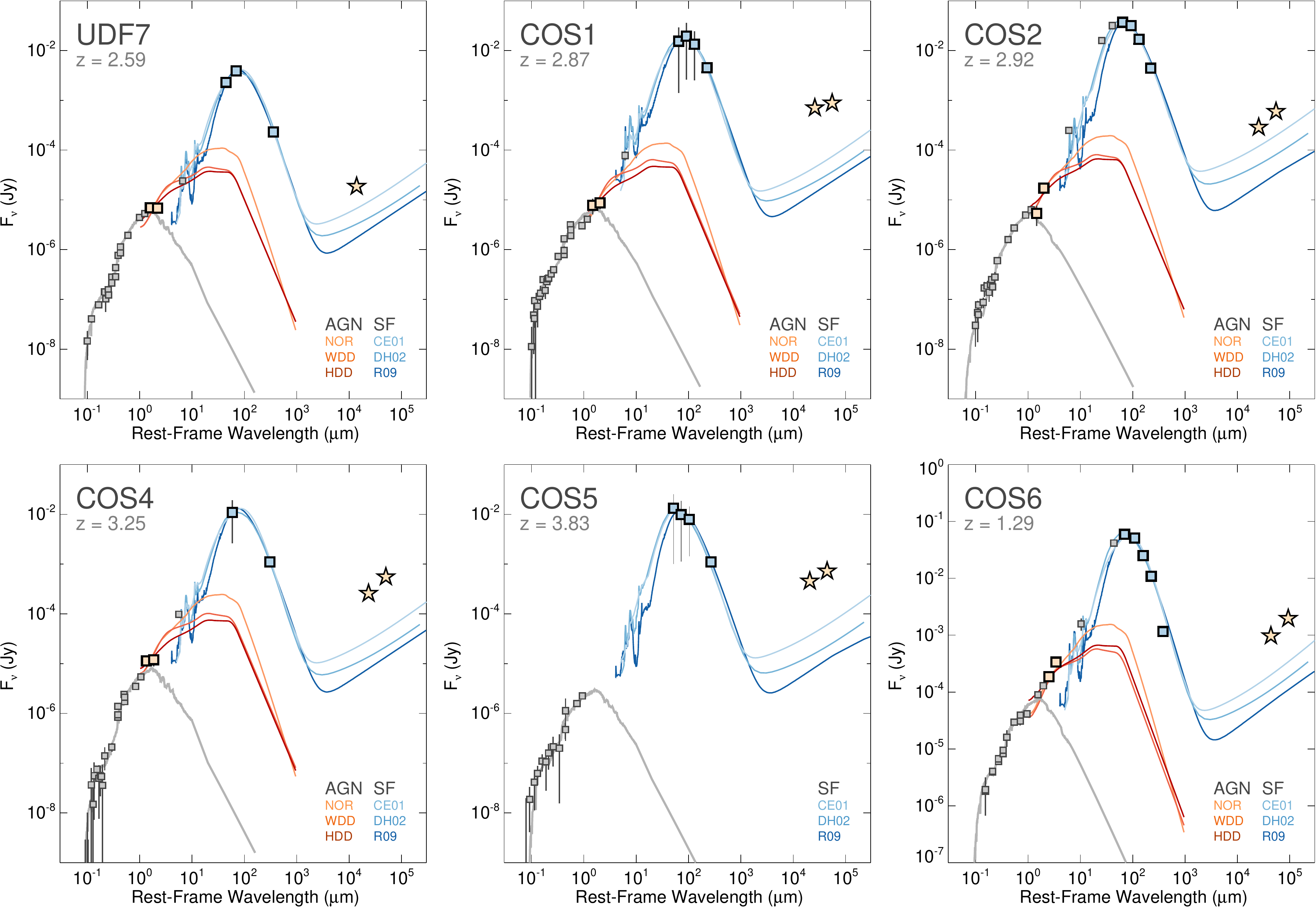}}
\caption{Spectral energy distributions (SEDs) of the radio-dominated AGNs. The blue squares are photometry longward of rest-frame 40 \micron\ used for the SFR estimation. Best-fit star-forming SEDs from the \citet{Rieke09}, \citet{DH02}, and \citet{CE01} libraries are labeled R09, DH02, and CE01, respectively. The red squares are the observed 5.8 and 8.0\,\micron\ photometry, to which we normalize the \citet{LyuRieke17} AGN SED templates to represent the maximal AGN emission; the normal, warm-dust-deficient, and hot-dust-deficient SEDs are indicated by NOR, WDD, and HDD, respectively. We note that COS5 is undetected at 5.8 and 8.0\,\micron, hence the absence of the maximal AGN SEDs. The grey lines are FAST \citep{Kriek09} best-fit stellar photospheric emission. Lastly, the stars show radio emission, which is clearly enhanced from the level associated with star-formation and is dominated by AGN emission. \label{fig:all_SED}}
\end{figure*}

Recently, the Atacama Large Millimeter/submillimeter Array (ALMA) now allows us to image the spatial distribution of the dust associated with star formation at sub-arcsecond resolution. The thermal continuum at, e.g., $870 - 1300$ \micron\ probes the rest-frame dust emission at $220 - 330$ \micron\ at $z = 3$, which is close to the peak of the spectral energy distribution (SED) of a typical dust-embedded star forming galaxy. At the same time, this emission is largely free from any AGN contribution, which typically plummets rapidly longward of 40 $\micron$ \citep{Elvis94, LyuRieke17}. Likewise, sub-arcsecond centimetric radio observations (at, e.g., $1-10$ GHz) from the Karl G. Jansky Very Large Array (VLA) can penetrate dust to trace synchrotron emission associated with the star formation, and is also sensitive to emission from any AGN core and jets. Any galaxy where the radio emission is enhanced to well above the level implied by the far-infrared/radio correlation for star formation \citep{Helou85} presents a robust radio AGN signature \citep{Donley05}, which we will refer hereafter as ``radio-dominated AGN''. In these cases, we can use the radio image to localize the SMBH in relation to the sub/millimeter morphologies of star formation.

In this Letter, we demonstrate the use of high resolution VLA and ALMA images to localize AGN in relation to the distribution of star formation in galaxies at $z \sim 3$. We adopt a $\Lambda$CDM cosmology with $\Omega_M = 0.3$, $\Omega_\Lambda = 0.7$, $H_{0} = 70~{\rm km\,s}^{-1}{\rm Mpc}^{-1}$, and the \citet{Chabrier03} IMF.

\begin{figure*}[ht]
\figurenum{2}
\centerline{\includegraphics[width=\textwidth]{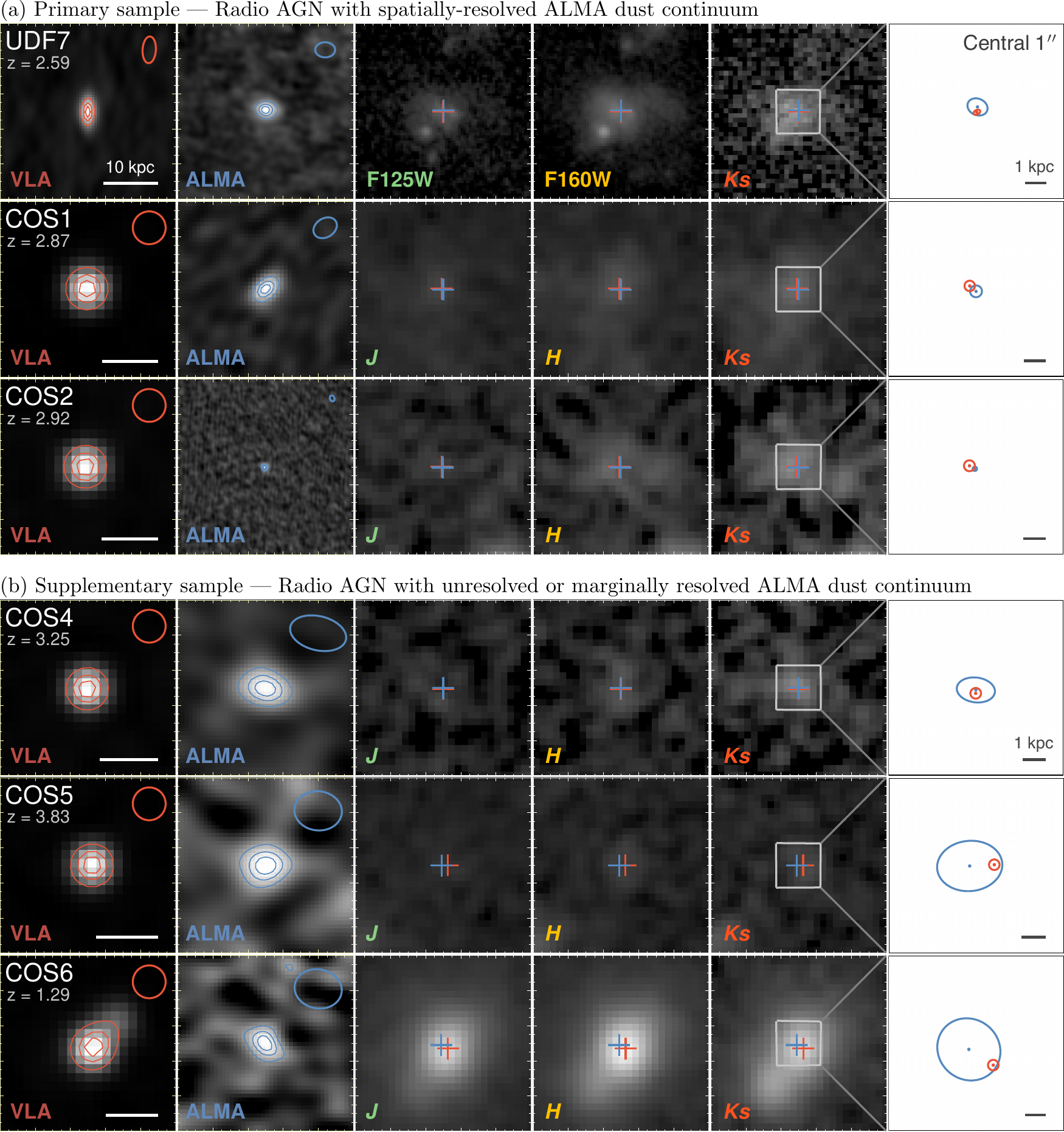}}
\caption{The six radio-dominated AGN with ALMA detections. From left to right are $4'' \times 4''$ image cutouts from (1) VLA (strongly AGN-dominated); (2) ALMA (nearly pure star formation); (3) 1.25\,\micron; (4) 1.6\,\micron; (5) 2.15\,\micron; and (6) a close-up schematic diagram of the central $1''$. The deep near-infrared images from VISTA and {\it HST} \citep{McCracken12, Ellis13} trace existing stellar mass distribution. VLA and ALMA synthesized beams are shown in their corresponding columns; the contours are 0.50, 0.75, and 0.90 of the peak flux; north is up, east is left. The red and blue crosses in columns $3-5$ indicate the centroid positions of AGN (VLA) and star-forming regions (ALMA), respectively. The ellipses in the close-up column indicate the $3\sigma$ positional uncertainty of the centroids of the star-forming and AGN emission (again, shown in blue and red). The only difference between the primary and the supplementary sample is that the ALMA spatial resolution of the former is sufficiently high to resolve the sub/millimeter star-forming region (Section \ref{sec:obs}). \label{fig:postage_stamps}}
\end{figure*}

\section{Sample Selection, Precise Estimation of Source Position and Size} \label{sec:obs}

To identify radio-dominated AGN, we extract sources from deep VLA images of the Hubble Ultra-Deep Field (HUDF) and COSMOS and compare their radio fluxes with the sub/millimeter fluxes from ALMA. Sources with radio fluxes $\geqslant0.7$ dex above the best fit far-infrared/radio correlation at the corresponding redshift (i.e., emitting in radio $\geqslant5\times$ more luminously than the level predicted by their star formation) are classified as radio-dominated AGN; the 0.7-dex threshold being $\approx3$ times the dispersion of the far-infrared/radio correlation \citep{Magnelli15}. This selection of radio AGN is inherently (and deliberately) conservative in the sense that it discriminates against AGNs with weakly enhanced radio emission, because our goal is to select a sample whose radio emission is strongly dominated by AGN to demonstrate accurate localization of the site of SMBH growth.

The HUDF and COSMOS VLA images \citep{Rujopakarn16, Smolcic17} have rms sensitivities of 0.3 and 2.3\,\uJybm\ at 6.0 and 3.0\,GHz, and synthesized beam sizes of $0\farcs3$ and $0\farcs75$, respectively. The \citet{Dunlop17} $2' \times 2'$ contiguous ALMA image of the HUDF has an rms sensitivity of 29\,\uJybm\ at 1.3\,mm (band 6) and a synthesized beam of $0\farcs4$. In COSMOS, we compiled pointed observations from programs ALMA\# 2011.0.00097.S, 2013.1.00034.S, and 2015.1.00137.S; these observations are at $0.87 - 1.3$\,mm \citep[bands 6 and 7;][]{Scoville17}. We re-reduce and re-image the ALMA observations with the CASA task {\tt CLEAN} and make the correction for the primary beam attenuation where necessary.

We extract VLA and ALMA sources down to $5\times$ the local rms noise using PyBDSM\footnote{http://www.astron.nl/citt/pybdsm} and cross match them with multiwavelength catalogs for photo-$z$ and stellar mass estimates based on $0.3-8$\,\micron\ SED fitting. In COSMOS, we use the \citet{Laigle16} catalog with a search radius of $0\farcs75$; multiwavelength catalog construction for the HUDF is described in \citet{Dunlop17}. We then construct the far-infrared/radio correlation as a function of redshift and identify sources with radio emission $\geqslant 0.7$ dex above this correlation. 

We identified five new radio-dominated AGN, in addition to UDF7, previously reported\footnote{Although UDF7 was identified from the \citet{Dunlop17} ALMA image, subsequent observations with higher resolution and signal-to-noise ratio from program ALMA\# 2013.1.01271.S (PI: Cibinel) is used in this study.} by \citet{Rujopakarn16}. The centroid position and deconvolved size of the VLA and ALMA emission were measured with 2D Gaussian fitting using the AIPS task {\tt JMFIT}; all Gaussian parameters were free. Sources with the {\tt JMFIT} nominal deconvolved sizes greater than zero in both axes are considered to be spatially resolved. Among the six AGNs, three were observed in sufficiently extended antenna configurations to spatially resolve their sub/millimeter emission (hereafter the ``primary sample''), which provides basic constraints on the size of their star-forming regions. The remaining three are observed at lower spatial resolutions (e.g., $\simeq1''$) that do not resolve the sub/millimeter emission. We refer to the latter three as the ``supplementary sample'' because future high resolution observations will be necessary to determine the location and morphology of their star-forming regions (e.g., distinguishing between disk-wide vs. nucleated). The sample is tabulated in Table \ref{tab:sample}; their SEDs are shown in Figure \ref{fig:all_SED}. 

Pinpointing the relative locations of star formation (which dominates the ALMA images) and AGN (which dominates the VLA images) depends on the accuracy of the centroid positions from 2D Gaussian fitting, $\sigma_{\rm pos}$, which relies primarily on (1) the absolute astrometric reference that is tied to the phase calibrator positions, which are accurate to 2 and 10 mas for VLA in the HUDF and COSMOS, respectively, and $\lesssim1$ mas for ALMA; and (2) the signal-to-noise ratio, SNR, of the detection and the synthesized beam size, $\theta_{\rm beam}$, which are related to positional uncertainties by $\sigma_{\rm pos} \approx \theta_{\rm beam}/(2 \times {\rm SNR})$ following \citet{Condon97}. We add these two uncertainties in quadrature for all positional uncertainties considered.

We verify the absolute astrometric calibration of the COSMOS VLA map by cross-matching the 7833 sources detected at $\geqslant 5\sigma$ with the {\it Gaia} DR1 catalog, which is tied to the International Celestial Reference Frame (ICRF) within 0.1 mas \citep{GaiaDR1}. This results in 81 matches. These overlapping sources indicated a median positional offset of $\Delta\alpha = 11$ mas and $\Delta\delta = 8$ mas between the VLA and {\it Gaia}, consistent with the 10 mas positional uncertainties of the {\it Gaia} DR1 for sources fainter than 11.5 mag \citep{Lindegren16}, see also \citet{Smolcic17} for a comparison with the Very Long Baseline Array data yielding consistent results. The absolute astrometry of the HUDF VLA map has previously been verified by \citet{Rujopakarn16}. For ALMA, the theoretical absolute astrometric accuracy at the frequency, baseline length, and signal-to-noise ratio for the observations of the primary sample is $8-26$ mas \citep{ALMAhandbook}, which is similar to those of the VLA in the HUDF and COSMOS. The astrometry in the supplementary sample is about $2 - 3$ times less precise. Therefore, the primary sample astrometry for both VLA and ALMA is sufficiently accurately tied to the ICRF to allow positional comparison at $\simeq30$ mas, corresponding to $\simeq 230$ pc at $z = 3$.

\begin{deluxetable*}{lccccccc}
\tablenum{2}
\tablecaption{Position and Size of AGN and Star-Forming Region}
\tablehead{\colhead{ID} & \colhead{$\sigma_{\rm pos}{\rm (AGN)}$} & \colhead{$\sigma_{\rm pos}{\rm (SF)}$} & \colhead{VLA Deconvolved} & \colhead{ALMA Deconvolved} & \colhead{$\Delta_{\rm p}$(SF, AGN)} & \colhead{$R_e$(SF)} & \colhead{Average \SFRSD}\\ 
\colhead{} & \colhead{(mas)} & \colhead{(mas)} & \colhead{FWHM ($''$)} & \colhead{FWHM ($''$)} & \colhead{(kpc)} & \colhead{(kpc)} & \colhead{(\Msunyrkpcsq)}
}
\startdata
UDF7 & $   4 \times   7$ & $ 20 \times  16$ &  $0.24^{+0.06}_{-0.07} \times 0.13^{+0.05}_{-0.13}$ & $0.35^{+0.11}_{-0.12} \times 0.23^{+0.15}_{-0.23}$ & $0.2 \pm 0.2$ & $1.1 \pm 0.5$ & $  20 \pm  10$ \\ 
COS1 & $  10 \times  10$ & $ 11 \times  11$ &  $0.11^{+0.02}_{-0.03}$ $\times < 0.02$ \phantom{...} & $0.38^{+0.08}_{-0.09} \times 0.12^{+0.11}_{-0.12}$ & $0.4 \pm 0.2$ & $0.8 \pm 0.4$ & $ 190 \pm 160$ \\ 
COS2 & $  10 \times  10$ & $  4 \times   4$ &  $0.14^{+0.04}_{-0.14} \times 0.10^{+0.08}_{-0.10}$ & $0.11^{+0.03}_{-0.04} \times 0.09^{+0.04}_{-0.05}$ & $0.3 \pm 0.1$ & $0.4 \pm 0.1$ & $1700 \pm 300$ \\ 
COS4$^*$ & $  10 \times  10$ & $ 37 \times  24$ &  $0.09^{+0.05}_{-0.09}$ $\times <0.06$ \phantom{...} & ... & $0.1 \pm 0.4$ & ... & ... \\ 
COS5$^*$ & $  10 \times  10$ & $ 63 \times  49$ &  $0.12^{+0.03}_{-0.04} \times 0.07^{+0.05}_{-0.07}$ & ... & $1.0 \pm 0.6$ & ... & ... \\ 
COS6$^*$ & $  10 \times  10$ & $ 63 \times  58$ &  $0.86^{+0.01}_{-0.01} \times 0.28^{+0.01}_{-0.01}$ & ... & $1.4 \pm 0.7$ & ... & ... 
\enddata
\tablecomments{Position and size are measured with the AIPS task {\tt JMFIT}. The $\sigma_{\rm pos}$ are quadratic sums of the positional uncertainties from the 2D Gaussian fit and that of the phase calibrator positions. The VLA and ALMA deconvolved size uncertainties are the `nominal', `minimum', and `maximum' size measures from {\tt JMFIT}; deconvolved size limits are $1\sigma$. $\Delta_{\rm p}$(SF, AGN) is the physical separation between the peak position of the star-forming and AGN emission. The asterisks in the ID column indicate objects in the supplementary sample, whose ALMA detections are unresolved, and hence the lack of $R_e$(SF) and \SFRSD\ measurements. \label{tab:result}}
\end{deluxetable*}

\begin{figure*}
\figurenum{3}
\centerline{\includegraphics[width=0.325\textwidth]{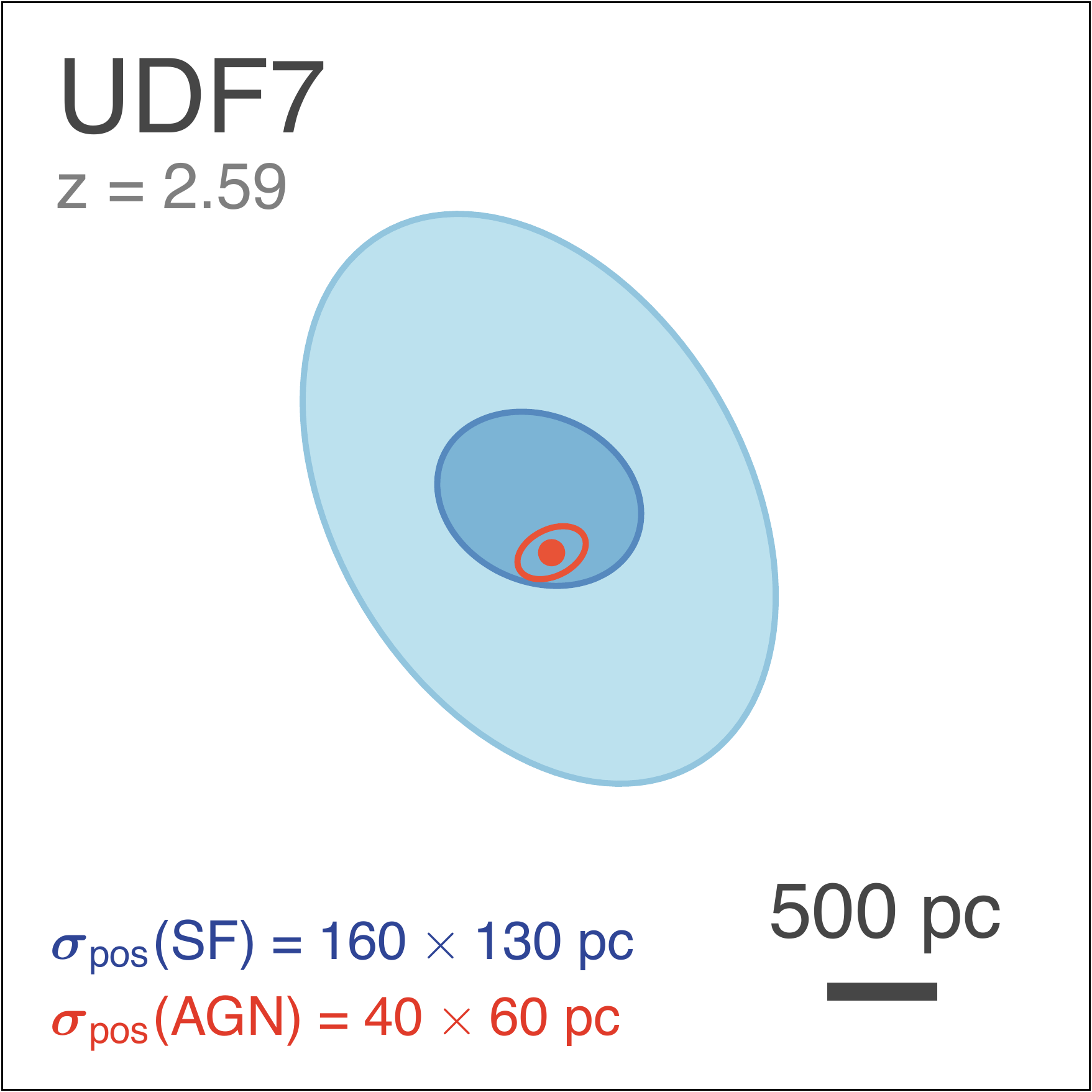}\phantom{--}\includegraphics[width=0.325\textwidth]{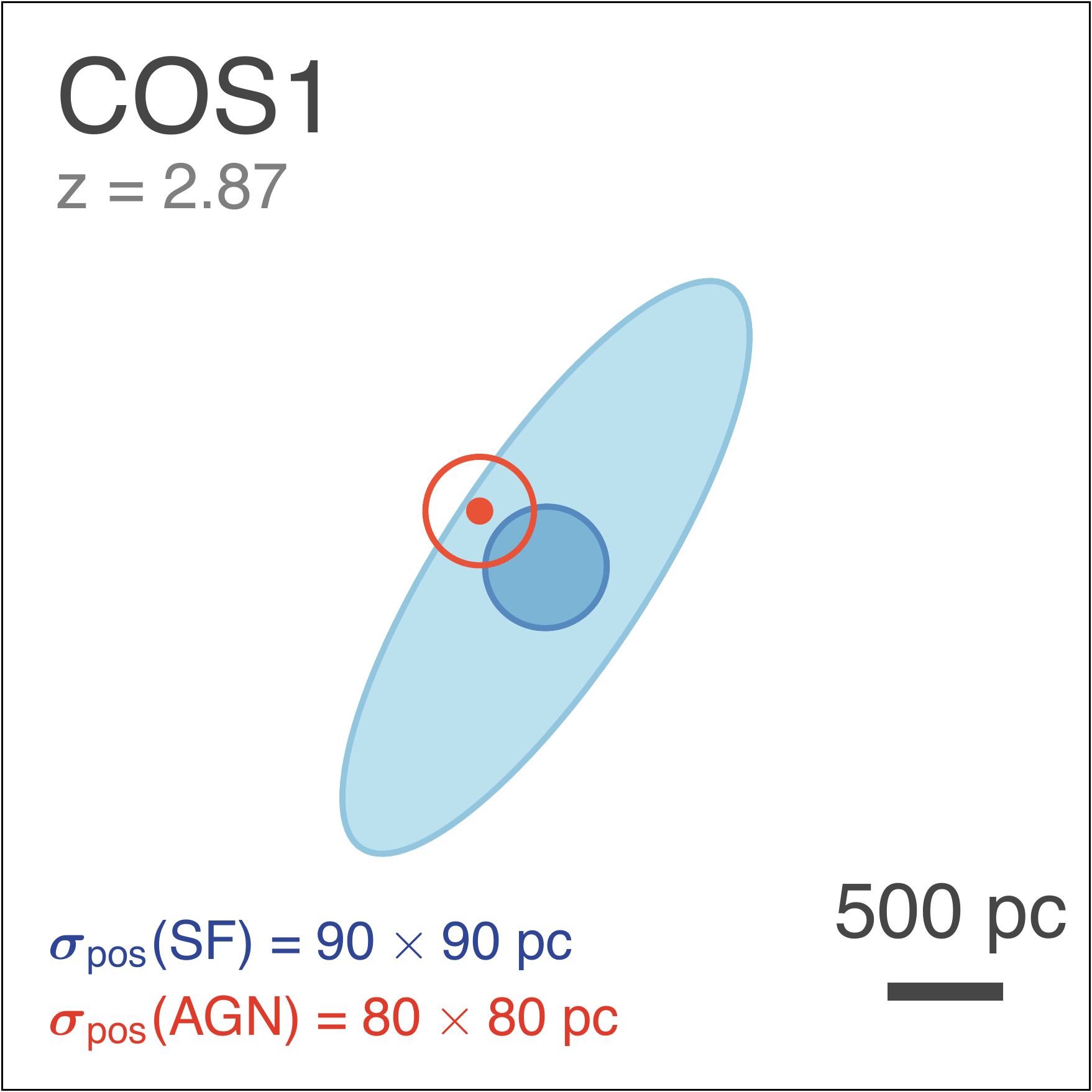}\phantom{--}\includegraphics[width=0.325\textwidth]{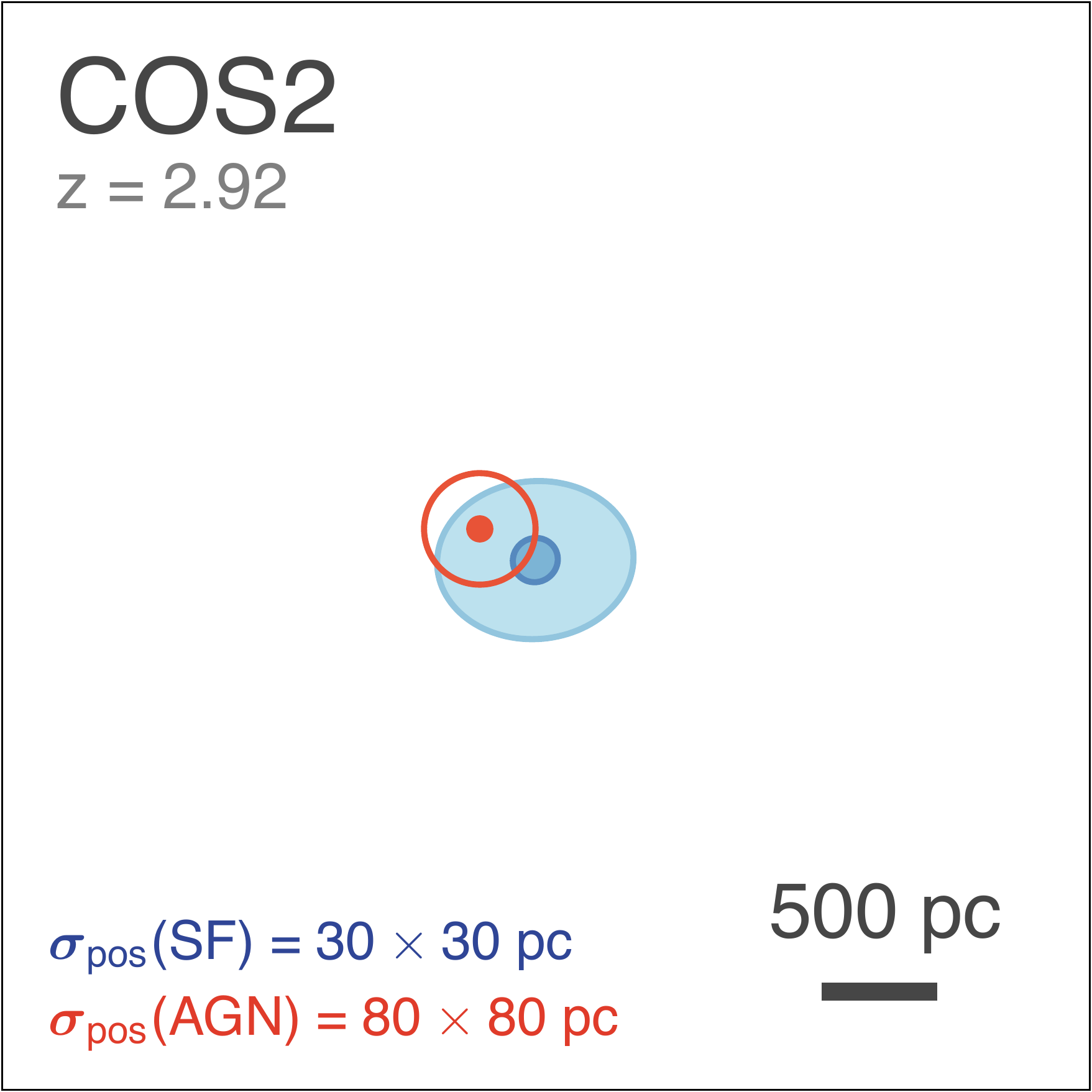}}
\caption{Schematic diagrams showing the central 5 kpc ($0\farcs6$) of AGNs in the primary sample; north is up, east is left. The red dots and red ellipses indicate the centroid positions and uncertainties, $3\,\sigma_{\rm pos}$(AGN), of the AGN emission. The dark and light blue ellipses are centroid uncertainties, $3\,\sigma_{\rm pos}$(SF), and the deconvolved size of the star-forming region, respectively. These data indicate that the AGN lie within the compact regions of intense star formation.\label{fig:close_up}}
\end{figure*}

\section{Results} \label{sec:results}

It is immediately apparent from the rightmost column of Figure \ref{fig:postage_stamps} that the star formation (ALMA) and AGN (VLA) are remarkably co-spatial. The AGN in the supplementary sample are also located within or at the edges of the uncertainty ellipses for the centroids of the star-forming regions. We report the positional uncertainties of the locations of star formation and AGN, their physical separations, and the sizes of the star-forming regions (for the primary sample) in Table \ref{tab:result}. 

\subsection{Spatial location of star formation, AGN, and stellar mass}\label{sec:results_spatialdistrib}

The high signal-to-noise ratio and small synthesized beam in our radio images allow us to pinpoint the centroids of AGN emission down to $\simeq30$ mas ($3\sigma$). The deconvolved sizes of the VLA emission have one or both axes consistent with being a point source at our resolution (Table \ref{tab:result}), indicating that their $\sim 10^{24} - 10^{26}$ W\,Hz$^{-1}$ radio power originates from very small regions. The small intrinsic sizes of these luminous radio sources \citep[cf. luminous parsec-scale core and jets, e.g.,][and references therein]{Zensus97} make them good tracers of the locations of SMBH growth.

For ALMA, the centroid uncertainties for the primary sample are larger, but they still allow localization of the region of star formation down to $13-200$ mas ($3\sigma$), corresponding to $0.1-1.5$ kpc at $z = 3$. These positional uncertainties are tabulated in Table \ref{tab:result} along with the physical separation between the centroids of the AGN and star-forming regions, $\Delta_{\rm p}$(SF, AGN), which are $0.2-0.4$ kpc for those in the primary sample and up to $\simeq$ 1.5 kpc in the supplementary one. For the primary sample, these separations are smaller than the sizes of the star-forming regions, which range from $R_{\rm e} = 0.4 \pm 0.1$ to $1.1 \pm 0.5$ kpc. That is, the AGN lies within the region of intense star formation (Figure \ref{fig:close_up}). Within the larger errors, this behavior is also the case for the supplementary sample.

The global (i.e., spatially-integrated) stellar masses are on average \logMstar\ $= 10.7$. However, no stellar mass concentration, indicative of a bulge, is detected in any of the near-infrared images (Figure \ref{fig:postage_stamps}; except for COS6, to be discussed below). Although their faint near-infrared detections are at locations consistent with those of the AGN and star-forming activities, higher resolution near-infrared imaging from, e.g., the {\it James Webb Space Telescope} will be required to determine conclusively whether AGN and star-forming activities are also located cospatially with the existing stellar mass concentrations. The lack of significant stellar mass build up at the location of the compact, intense star formation would be consistent with an early phase of ongoing bulge assembly.  

It is worth highlighting that COS6 is different from the rest of the sample: it has the largest stellar mass, \logMstar\ $= 11.6$, lowest gas fraction, $f_{\rm gas} = 0.12$ (Section \ref{sec:results_SFR}), lowest redshift, $z = 1.29$, and has an asymmetric radio morphology that extends beyond the optical extent, suggesting that it has a radio jet. These characteristics suggest that COS6 may be ending the phase of intense star formation and is transitioning to become a radio-mode AGN \citep{Croton06} as is typically found in local massive ellipticals \citep{Brown11}.

\subsection{Dust mass, gas fraction, and star formation rate}\label{sec:results_SFR}

We estimate the dust and gas masses using the same approach as in \citet{Rujopakarn16} and assuming a $T_{\rm dust}$ of 25 K \citep{Scoville17}. The uncertainties are estimated by deriving masses for $T_{\rm dust}$ ranging over  $20-30$ K (for the maximum and minimum estimates, respectively). We use a \citet{LiDraine01} dust mass absorption coefficient and a gas-to-dust ratio of 100 \citep{Leroy11, Magdis12}. The results are tabulated in Table \ref{tab:sample}. With exception of COS6, these AGN hosts are gas-rich systems consistent with being SFGs undergoing rapid assembly of their stellar masses. 

We estimate the SFR of these galaxies by fitting libraries of infrared SED templates of star-forming galaxies to the far-infrared observations to estimate the infrared luminosity, \LIR, and convert \LIR\ to SFR using the \citet{Kennicutt98} conversion with a correction to the \citet{Chabrier03} IMF. The \citet{Rieke09} SED library is adopted for \LIR\ estimation, with the best-fit SEDs from the \citet{CE01} and \citet{DH02} SED libraries shown for comparison in Figure \ref{fig:all_SED}. To mitigate the possible contamination from AGN, we use only observations at rest-frame wavelengths longward of 40 \micron\ for the fit, i.e. from ALMA and {\it Herschel} \citep[e.g.,][]{Lutz11, Oliver12}. We note that the radio luminosity of the sample would indicate SFRs of order $10^3 - 10^4$ \Msunyr\ if all radio emission were of star formation origin, assuming the \citet{Bell03} radio SFR indicator. Comparing the radio-implied SFR with the fiducial, far-infrared estimates indicates that only $2-15$\% (average 8\%) of the radio emission is powered by star formation, independently confirming that the compact radio emission unambiguously localizes the AGN.

To independently verify the level of AGN contribution to \LIR, we normalize the \citet{LyuRieke17} AGN SED templates to the {\it Spitzer}/IRAC 5.8 and 8.0 \micron\ photometry, which is the spectral region where significant contribution from AGN is expected \citep{AH06}, and estimate the \LIR\ associated to these templates (Figure \ref{fig:all_SED}). This exercise assumes that all of the $5.8 - 8.0$ \micron\ emission is of AGN origin, thereby representing a maximal AGN emission scenario. Yet we find that the maximum AGN contribution to \LIR\ is $\simeq 2-11\%$, further assuring that our far-infrared SFR estimates are robust against AGN contamination.

\section{Discussion} \label{sec:conclusion}

This work demonstrates the unique complementarity of VLA and ALMA observations to pinpoint the relative sites of both the AGN and star formation in galaxies at the peak epoch of galaxy assembly. While the sample compiled here is by no means homogeneous (their sub/millimeter and radio luminosities span more than an order of magnitude), ALMA can be used to construct a uniformly-selected sample of SFGs hosting radio-dominated AGNs by conducting sufficiently sensitive sub/millimeter observations of $z \sim 1-3$ radio AGN candidates. An obvious starting point is to conduct ALMA observations of AGN candidates among the $\simeq150$ sources in the \citet{Smolcic17} COSMOS image at $z > 1.5 $ that are detected by the VLA at $>40\sigma$ at a spatial resolution comparable to those of our primary sample to reach the SFR sensitivities of the level of the main-sequence of SFGs. This level of VLA signal-to-noise ratio will allow $\lesssim 10$ mas localization of the AGN.

The results of this work show the potential of such a sample: we have localized the sites of SMBH growth in three galaxies at $z \sim 3$ down to $\lesssim100$ pc (1$\sigma$ positional uncertainty), and in relation to the surrounding regions of intense star formation. For these three galaxies, whose ALMA-detected dust continuum is spatially resolved, the AGNs are found within the compact, gas-rich regions of intense star formation (Figure \ref{fig:close_up}).

The star forming regions in these galaxies are compact, with $R_e = 0.4-1.1$ kpc, implying a \SFRSD\ of $20-1700$ \Msunyrkpcsq, markedly higher than those of main sequence SFGs at $z \sim 2$ \citep{Rujopakarn16}, and are more in line with submillimeter galaxies \citep{Simpson15, Ikarashi17} and the nuclei of bulge-forming SFGs \citep{Barro16}. This level of \SFRSD\ will likely drive outflows \citep{Newman12, Bordoloi14}. If the star-formation and AGN-driven outflows \citep[e.g.,][]{Mullaney13} do not completely disrupt the cold gas supply, and both types of activity proceed for the duration of their typical stellar mass doubling time of $\simeq 0.2$ Gyr, then the newly formed stellar mass within the central $\sim$kpc will be of order $10^{11}$ \Msun. Likewise, if the SMBH is allowed to accrete at a rate predicted by the correlation between SFR and the average SMBH accretion rate relation \citep{Chen13} for the same duration of 0.2 Gyr, the accreted SMBH mass will be $10^{6.7} - 10^{7.8}$ \Msun, similar to those found in local massive galactic bulges. That is, the ongoing episode of star formation and SMBH growth in these galaxies is potentially capable of producing bulge stellar masses and SMBH masses on the local scaling relations. This possibility, of course, depends on the yet-to-be characterized star-formation- and AGN-driven outflows. While these vital details are still missing, we have demonstrated that AGN and star formation in these systems are cospatial and therefore are likely being fed with a common supply of cold gas. This is consistent with a picture of {\it in-situ} bulge assembly that proceeds concurrently and cospatially with SMBH growth.\\

W.R. acknowledges support from the JSPS KAKENHI Grant Number JP15K17604, Thailand Research Fund/Office of the Higher Education Commission Grant Number MRG6080294, and Chulalongkorn University's Ratchadapiseksompot Endowment Fund. R.J.I. acknowledges support from the European Research Council through the Advanced Grant COSMICISM 321302. V.S. acknowledges support from the European Union's Seventh Frame-work program under grant agreement 337595 (ERC Starting Grant, ``CoSMass''). This paper makes use of the following ALMA data: ADS/JAO.ALMA\#, 2011.0.00097.S, 2013.1.00034.S, 2013.1.01271.S, and 2015.1.00137.S. ALMA is a partnership of ESO (representing its member states), NSF (USA) and NINS (Japan), together with NRC (Canada) and NSC and ASIAA (Taiwan) and KASI (Republic of Korea), in cooperation with the Republic of Chile. The Joint ALMA Observatory is operated by ESO, AUI/NRAO and NAOJ. This work was supported by the World Premier International Research Center Initiative (WPI), MEXT, Japan.



\end{document}